\documentclass[aps,preprint]{revtex4-1}
\usepackage{bm}
\usepackage{amsfonts}
\usepackage{amsmath}
\usepackage{amssymb}
\usepackage{graphicx}

\begin{document}

\title{Comment on "The low lying modes of triplet-condensed neutron matter
and their effective theory"}
\author{L. B. Leinson}

\affiliation{Institute of Terrestrial Magnetism, Ionosphere 
and Radio Wave Propagation
RAS (IZMIRAN), 142190 Troitsk, Moscow, Russia}

\begin{abstract}
Contrary to what is claimed in the article by P. F. Bedaque and A. N.
Nicholson [Phys. Rev. C 87, 055807 (2013)], their result do not contradict
but rather complement the conclusion of the paper by L. B. Leinson [Phys.
Rev. C 85, 0655021 (2012)] with respect to the low lying nonunitary
excitations in the anisotropic neutron superfluid. In fact, the
corresponding Goldstone modes (angulons), existing at zero temperature, can
be found from Eq. (59) of the work by Leinson. Being massless at absolute
zero these modes acquire a mass gap at finite temperature.
\end{abstract}

\maketitle



The authors of a recent paper \cite{Bedaque} derived a low-energy effective
theory describing the massless Goldstone bosons (angulons) associated with
broken rotational symmetry in a $^{3}$P$_{2}\left( m_{j}=0\right) $
condensed neutron superfluid at zero temperature. The authors claim that
their result is in contradiction with the work by Leinson \cite{L12}. Below,
we argue that this is not the case and the massless Goldstone modes are the
solutions of the general dispersion equation at $T=0$.

In Ref. \cite{L12}, the author studied the collective excitations of the $%
^{3}$P$_{2}(m_{j}=0)$ condensate at arbitrary temperatures. It was found
that the eigenmodes of the pseudo-vector current should satisfy Eq. (59) of
this work. The dispersion equation has been solved for the case of vanishing
wave vector, $q=0$, and the conclusion was made that, \emph{at finite
temperature}, there are no gapless modes associated with broken rotational
symmetry. Instead, there have been found the nonunitary excitations
corresponding to periodic flapping of the total angular momentum $\mathbf{j}$
(the same angular oscillations) which are gapless at zero temperature but
having a mass gap, $\tilde{\omega}_{1}$, at finite temperatures below the
critical temperature $T_{c}$.

This however is not in a contradiction to the results of Ref. \cite{Bedaque}
because, in the case of anisotropic pairing, the Nambu-Goldstone theorem,
valid at absolute zero, is violated at finite temperatures. To be convinced
in the violation of the Goldstone theorem one needs to solve the equation
(59) of Ref. \cite{L12} for finite values of the temperature and wave
vector. We solve this equation for the case when the temperature $T$ is very
small as compared to the superfluid energy gap $\Delta $, and $\omega
,qV_{F}\ll 2\Delta $ (System of units $\hbar =c=1$, and the Boltzmann
constant $k_{B}=1$). The Fermi velocity is small in the nonrelativistic
system of neutrons, $V_{F}\ll 1$.

At low temperature, when  $y\equiv \Delta \left( T\right) /T\gg 1$, in Eq.
(59) of Ref. \cite{L12},\ we obtain $A\simeq \ln \pi -\ln \left( \gamma \bar{%
b}y\right) -2\varkappa \exp \left( -y\bar{b}\right) $, and%
\begin{equation}
\mathcal{I}_{FF}\simeq \frac{1}{2\bar{b}^{2}}\left( 1+\delta _{FF}\right) ,\
\ \delta _{FF}\equiv \frac{1}{s^{2}}\left( \cos ^{2}\theta _{\mathbf{nq}%
}-s^{2}\right) \sqrt{\frac{2\pi }{\bar{b}y}}e^{-y\bar{b}},  \label{FFT}
\end{equation}%
where $s=\omega /\left( qV_{F}\right) $, and $\theta _{\mathbf{qn}}$  is the
angle between the wave vector and the direction of the quasiparticle
momentum.

In the case $\sqrt{y}\gg 1$ one can neglect $\delta _{FF}$, thus obtaining
the following dispersion equation for the oscillations with $m_{j}=\pm 1$:  
\begin{equation}
\left\langle \left( s^{2}-\cos ^{2}\theta _{\mathbf{qn}}\right) \frac{%
\mathbf{b}_{1}^{\ast }\mathbf{b}_{1}}{\bar{b}^{2}}-\frac{2\varkappa \Delta
^{2}}{q^{2}V_{F}^{2}}\mathbf{b}_{1}^{\ast }\mathbf{b}_{1}e^{-y\bar{b}%
}\right\rangle ^{2}-\left\langle \frac{\mathbf{b}_{1}\mathbf{b}_{1}}{\bar{b}%
^{2}}\cos ^{2}\theta _{\mathbf{qn}}\right\rangle \left\langle \frac{\mathbf{b%
}_{1}^{\ast }\mathbf{b}_{1}^{\ast }}{\bar{b}^{2}}\cos ^{2}\theta _{\mathbf{qn%
}}\right\rangle =0,  \label{dEq}
\end{equation}%
where the angle brackets denote the averaging over directions of the
quasiparticle momentum,%
\begin{equation}
\varkappa \equiv \frac{1}{2}e^{1/4}K_{0}\left( 1/4\right) \simeq 0.989667,
\label{kappa}
\end{equation}%
and $K_{0}\left( z\right) $ being the Bessel function. The vectors $\mathbf{b%
}_{1}$ and $\mathbf{\bar{b}}$ are defined in Ref. \cite{L12}: 
\begin{equation*}
\bar{b}^{2}=\frac{1}{2}\left( 1+3\cos ^{2}\theta \right) ,\ \ \mathbf{b}%
_{1}^{\ast }\mathbf{b}_{1}=\frac{3}{4}\left( 1+\cos ^{2}\theta \right).
\end{equation*}%
Here $\theta $ is the polar angle on the Fermi surface. 

Performing the angular integration one can find the solutions in the form: 
\begin{equation}
\omega ^{\left( 1\right) }=\sqrt{\frac{9}{4\pi \sqrt{3}+18}q_{\perp
}^{2}V_{F}^{2}+\frac{27-2\pi \sqrt{3}}{6\pi \sqrt{3}+27}q_{z}^{2}V_{F}^{2}+%
\tilde{\omega}_{1}^{2}\left( T\ll \Delta \right) },  \label{w1T}
\end{equation}%
\begin{equation}
\omega ^{\left( 2\right) }=\sqrt{\frac{16\pi \sqrt{3}-27}{12\pi \sqrt{3}+54}%
q_{\perp }^{2}V_{F}^{2}+\frac{\left( 27-2\pi \sqrt{3}\right) }{6\pi \sqrt{3}%
+27}q_{z}^{2}V_{F}^{2}+\tilde{\omega}_{1}^{2}\left( T\ll \Delta \right) },
\label{w2T}
\end{equation}%
where%
\begin{equation}
\tilde{\omega}_{1}\left( T\ll \Delta \right) =\frac{6\varkappa ^{1/2}}{\sqrt{%
2\pi \sqrt{3}+9}}\Delta \left( T\right) \left\langle \mathbf{b}_{1}^{\ast }%
\mathbf{b}_{1}e^{-y\bar{b}}\right\rangle ^{1/2}.  \label{wgap}
\end{equation}

In the limit $T=0$, from Eqs. (\ref{w1T}) and (\ref{w2T}) we obtain the
massless solutions which numerically recover the dispersion law for the
angulons in Eq. (25) of Ref. \cite{Bedaque}. In particular case where the
propagation is along the $z$-axis and in the orthogonal direction the
corresponding velocities $s_{i}\equiv \omega _{i}/\left( qV_{F}\right) $
(for two modes 1 and 2) are 
\begin{equation}
s_{z}^{\left( 1,2\right) }=\sqrt{\frac{27-2\pi \sqrt{3}}{6\pi \sqrt{3}+27}}%
=0.519\,8,  \label{sz0}
\end{equation}%
\begin{equation}
s_{\perp }^{\left( 1\right) }=\sqrt{\frac{9}{4\pi \sqrt{3}+18}}=0.475\,7,
\label{syx0}
\end{equation}%
\begin{equation}
s_{\perp }^{\left( 2\right) }=\sqrt{\frac{16\pi \sqrt{3}-27}{12\pi \sqrt{3}%
+54}}=0.709\,6.  \label{sxy0}
\end{equation}%
Thus, for \emph{\ zero temperature} we find two Goldstone bosons which are
associated with the breaking of rotational invariance in two planes, in
accordance with Ref. \cite{Bedaque}.

\begin{figure}[h]
\includegraphics{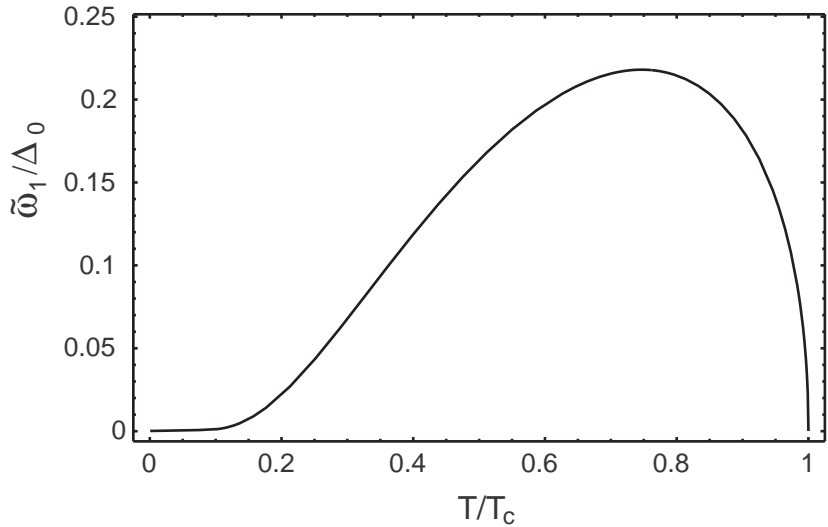}
\caption{The mass gap of the low lying nonunitary excitations versus reduced
temperature in the neutron $^{3}P_{2}(m_{j}=0)$ condensate. The gap energy
is given in units of $\Delta _{0}\equiv \Delta \left( T=0\right) $.}
\label{fig1}
\end{figure}

At finite temperature the nonunitary Goldstone bosons acquire a mass gap, $%
\tilde{\omega}_{1}$, which is exponentially small at low temperatures but
considerably grows when the temperature increases. Its temperature
dependence is presented in Fig. \ref{fig1} which demonstrates the result
obtained in Ref. \cite{L12}. The mass gap $\tilde{\omega}_{1}$ passes
through a maximum and tends to zero both for $T\rightarrow 0$ and for $%
T\rightarrow T_{c}$. 

Thus the Goldstone theorem is violated in the case of spontaneous breaking
of the rotation symmetry at finite temperatures. A natural explanation of
this phenomenon is as follows. In our model, the low lying oscillations are
associated with a flapping of the total angular momentum. This can be
imagined as a departure of the symmetry axis of the bound pair about the
symmetry axis of the equilibrated condensate, what is equivalent to
oscillations of the preferred direction of the Cooper pair relative to the
axis of the energy gap in the quasiparticle energy. The quasiparticle
readjustment can not follow this rapid motion. As a result any rotation of
the total angular momentum from its original orientation will cost energy
which grows with the number of broken pairs and with the magnitude of the
superfluid energy gap.

Immediately below the critical temperature the moment of inertia, associated
with the quasiparticles, increases along with increase of the superfluid
energy gap at lowering of the temperature. However the moment of inertia
declines owing to a decrease of number of thermal quasiparticles along with
the further lowering of the temperature. At zero temperature the restoring
force is expected to vanish along with the number of thermal quasiparticles.
Accordingly, we obtain the temperature behavior of the function $\tilde{%
\omega}_{1}$ shown in the above plot.

The mass gap in the angulon spectra leads to important consequences because,
unlike the massless modes predicted in Ref. \cite{Bedaque}, the gapped waves
with $\omega >qV_{F}$ do not undergo the Landau damping and can propagate at 
finite temperature.

It should be noted that a possible violation of the Nambu-Goldstone theorem
at finite temperatures is known (See e.g. \cite{W}).

\end{document}